\documentclass{ws-procs9x6}
\usepackage{graphicx}
\usepackage[dvips]{epsfig}
\usepackage{amssymb}
\def\sigup{\sigma^{\uparrow\uparrow}}
\def\sigdw{\sigma^{\uparrow\downarrow}}
\def\gsim{\mathrel{\rlap{\lower4pt\hbox{\hskip1pt$\sim$}}\raise1pt\hbox{$>$}}}
\def\Att{A_{TT}}

\def\Journal#1#2#3#4{{#1} {\bf #2}, #3 (#4)}
\def\NCA{\em Nuovo Cimento}

\def\PLB{{\em Phys. Lett.}  B}
\def\PRL{\em Phys. Rev. Lett.}
\def\PR{\em Phys. Rep.}
\def\PRD{{\em Phys. Rev.} D}
\def\PRE{{\em Phys. Rev.} E}

\def\EPJC{{\em Eur. Phys.} J}

\newcommand{\be}{\begin{equation}}
\newcommand{\ee}{\end{equation}}
\newcommand{\ba}{\begin{eqnarray}}
\newcommand{\ea}{\end{eqnarray}}

\begin{document}

\title{SPIN Physics at GSI\footnote{\uppercase{A} short version of this report
    can be found in Ref.~$^1$.}}

\author{Frank Rathmann}

\address{
Institut f\"ur Kernphysik\\
Forschungszentrum J\"ulich\\
52425 J\"ulich, Germany\\ 
E-mail: f.rathmann@fz-juelich.de
}

\maketitle

\abstracts{ Polarized antiprotons produced by spin filtering with an
  internal polarized gas target provide access to a wealth of single--
  and double--spin observables, thereby opening a window to physics
  uniquely accessible with the HESR at FAIR.  This includes a first
  measurement of the transversity distribution of the valence quarks
  in the proton, a test of the predicted opposite sign of the
  Sivers--function, related to the quark distribution inside a
  transversely polarized nucleon, in Drell--Yan (DY) as compared to
  semi--inclusive DIS, and a first measurement of the moduli and the
  relative phase of the time--like electric and magnetic form factors
  $G_{E,M}$ of the proton.  In polarized and unpolarized $p\bar{p}$
  elastic scattering open questions like the contribution from the odd
  charge--symmetry Landshoff--mechanism at large $|t|$ and
  spin--effects in the extraction of the forward scattering amplitude
  at low $|t|$ can be addressed.  }

%%%%%%%%%%%%%%%%%%%%%%%%%%%%%%%%%%%%%%%%%%%%%%%%%%%%%%%%%%%%%%%%%%%%%%%%%%%%%
\section{Physics Case}
%%%%%%%%%%%%%%%%%%%%%%%%%%%%%%%%%%%%%%%%%%%%%%%%%%%%%%%%%%%%%%%%%%%%%%%%%%%%%

The polarized antiproton--proton interactions at the High Energy
Storage Ring (HESR) at the future Facility for Antiproton and Ion
Research (FAIR) will provide unique access to a number of new
fundamental physics observables, which can be studied neither at other
facilities nor at HESR without transverse polarization of protons and
antiprotons.
  
\subsection{The transversity distribution} is the last leading--twist missing
piece of the QCD description of the partonic structure of the nucleon.
It describes the quark transverse polarization inside a transversely
polarized proton \cite{bdr}. Unlike the more conventional unpolarized
quark distribution $q(x,Q^2)$ and the helicity distribution $\Delta
q(x,Q^2)$, the transversity $h^q_1(x,Q^2)$ can neither be accessed in
inclusive deep--inelastic scattering of leptons off nucleons nor can
it be reconstructed from the knowledge of $q(x,Q^2)$ and $\Delta
q(x,Q^2)$.  It may contribute to some single--spin observables, but
always coupled to other unknown functions.  The transversity
distribution is directly accessible uniquely via the {\bf double
  transverse spin asymmetry} $A_{TT}$ in the Drell--Yan production of
lepton pairs. The theoretical expectations for $A_{TT}$ in the
Drell--Yan process with transversely polarized antiprotons interacting
with a transversely polarized proton target at HESR are in the
0.3--0.4 range \cite{jpsi,efremov}; with the expected beam
polarization achieved using a dedicated low--energy antiproton
polarizer ring (AP) of $P\approx 0.3$ and the luminosity of HESR, the
PAX experiment\footnote{\uppercase{PAX} collaboration (\uppercase{{\bf
      P}}olarized \uppercase{{\bf A}}ntiproton \uppercase{E{\bf
      X}}periments). For the web--site, see
  http://www.fz-juelich.de/ikp/pax.} is uniquely suited for the
definitive observation of $h^q_1(x,Q^2)$ of the proton for the valence
quarks.  The determination of $h^q_1(x,Q^2)$ will open new pathways to
the QCD interpretation of single--spin asymmetry (SSA) measurements.
In conjunction with the data on SSA from the HERMES collaboration
\cite{hermesSSA}, the PAX measurements of the SSA in Drell--Yan
production on polarized protons can for the first time provide a test
of the theoretical prediction \cite{Collins} of the reversal of the
sign of the Sivers function \cite{SiversFunction} from semi--inclusive
DIS to Drell--Yan production.
  
\subsection{Magnetic and electric form factors} The origin of the 
unexpected $Q^2$--dependence of the ratio of the magnetic and electric
form factors of the proton as observed at the Jefferson laboratory
\cite{perdrisat} can be clarified by a measurement of their relative
phase in the time--like region, which discriminates strongly between
the models for the form factor. This phase can be measured via SSA in
the annihilation $ \bar{p} p^{\uparrow} \to e^+e^-$ on a transversely
polarized target \cite{d,brodsky}. The first ever measurement of this
phase at PAX will also contribute to the understanding of the onset of
the pQCD asymptotics in the time--like region and will serve as a
stringent test of dispersion theory approaches to the relationship
between the space--like and time--like form factors
\cite{Geshkenbein74,HMDtimelike,egle}.  The double--spin asymmetry
will allow independently the $G_E-G_M$ separation and serve as a check
of the Rosenbluth separation in the time--like region which has not
been carried out so far.
  
\subsection{Hard scattering} Arguably, in $p\bar{p}$ elastic scattering 
the hard scattering mechanism can be checked beyond $|t| =
\frac{1}{2}(s-4m_p^2)$ accessible in the $t$--$u$--symmetric $pp$
scattering, because in the $p\bar{p}$ case the $u$--channel exchange
contribution can only originate from the strongly suppressed exotic
dibaryon exchange.  Consequently, in the $p\bar{p}$ case the hard
mechanisms \cite{Matveev,BrodskyFarrar,KrollElastic} can be tested at
$t$ almost twice as large as in $pp$ scattering. Even unpolarized
large angle $p\bar{p}$ scattering data can shed light on the origin of
the intriguing oscillations around the $s^{-10}$ behavior of the
$90^0$ scattering cross section in the $pp$ channel and put stringent
constraints on the much disputed odd--charge conjugation Landshoff
mechanism \cite{Landshoff,RalstonPire,Ramsey,DuttaGao}.  If the
Landshoff mechanism is suppressed then the double transverse asymmetry
in $p\bar{p}$ scattering is expected to be as large as the one
observed in the $pp$ case.

\section{Towards an asymmetric polarized antiproton--proton collider at FAIR}
The possibility to test the nucleon structure via double spin
asymmetries in polarized proton--antiproton reactions at the HESR ring
of FAIR at GSI has been suggested by the PAX collaboration in 2004
\cite{paxloi}.  Since then, there has been much progress, both in
understanding the physics potential of such an
experiment~\cite{jpsi,efremov,brodsky_new,zavada} and in studying the
feasibility of efficiently producing polarized antiprotons~\cite{ap}.
The physics program of such a facility would extend to a new domain
the exceptionally fruitful studies of the nucleon structure performed
in unpolarized and polarized deep inelastic scattering (DIS), which
have been at the center of high energy physics during the past four
decades. As mentioned earlier, a direct measurement of the
transversity distribution function $h_1^q(x,Q^2)$, one of the last
missing fundamental pieces in the QCD description of the nucleon, is
unique. In the available kinematic domain of the proposed experiment,
which covers the valence region, the Drell--Yan double transverse spin
asymmetry was recently predicted to be as large as
0.3~\cite{jpsi,efremov}.  Other novel tests of QCD at such a facility
include the polarized elastic hard scattering of antiprotons on
protons and the measurement of the phases of the time--like form
factors of the proton (see Ref.~\cite{paxloi}).  A viable practical
scheme\footnote{The basic approach to polarizing and storing
  antiprotons at HESR--FAIR is based on solid QED calculations of the
  spin transfer from electrons to antiprotons \cite{HOMeyer}, which is
  being routinely used at Jefferson Laboratory for the electromagnetic
  form factor separation \cite{JlabFF}, and which has been tested and
  confirmed experimentally in the FILTEX experiment~\cite{Filtex}.}
which allows us to reach a polarization of the stored antiprotons at
HESR--FAIR of $\simeq 0.3$ has been worked out and published in
Ref.~\cite{ap}.

\begin{figure}[h]
\centerline{\includegraphics[width=\linewidth]{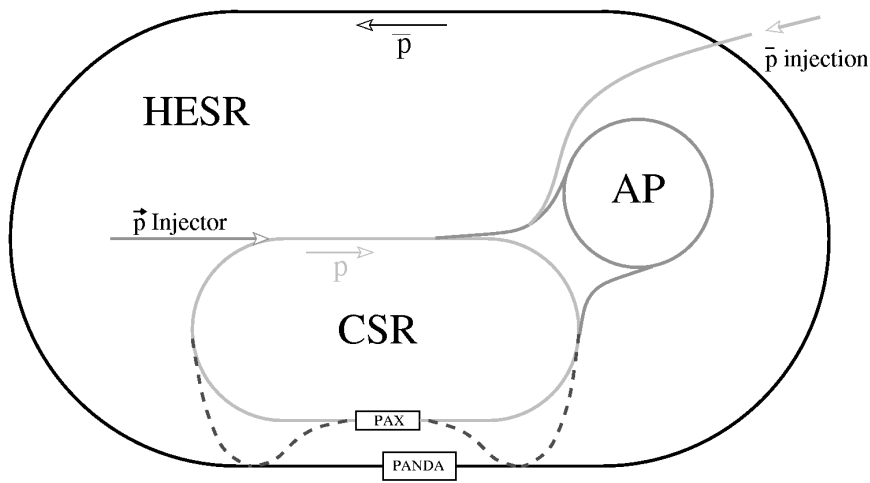}}
\caption{\small The proposed accelerator set--up at the HESR (black), 
  with the equipment used by the PAX collaboration in Phase I: CSR
  (light gray), AP, beam transfer lines and polarized proton injector
  (dark gray). In Phase II, by adding two transfer lines (dashed), an
  asymmetric collider is set up. It should be noted that, in this
  phase, also fixed target operation at PAX is possible. }
\label{CSRring}
\end{figure}

The PAX Letter--of--Intent was submitted on January 15, 2004. The
physics program of PAX has been positively reviewed by the QCD Program
Advisory Committee (PAC) on May 14--16, 2004 \cite{paxweb}.  The
proposal by the ASSIA collaboration \cite{ASSIA} to utilize a
polarized solid target and bombard it with a 45~GeV unpolarized
antiproton beam extracted from the synchrotron SIS100 has been
rejected by the GSI management.  Such measurements would not allow one
to determine $h^q_1(x,Q^2)$, because in single spin measurements
$h^q_1(x,Q^2)$ appears always coupled to another unknown fragmentation
function.  Following the QCD--PAC report and the recommendation of the
Chairman of the committee on Scientific and Technological Issues (STI)
\cite{paxweb} and the FAIR project coordinator, the PAX collaboration
has optimized the technique to achieve a sizable antiproton
polarization and the proposal for experiments at GSI with polarized
antiprotons \cite{ap}.  From various working group meetings of the PAX
collaboration, presented in part in 2004 at several workshops and
conferences \cite{paxweb}, we conclude:
%maybe we should quote here the PAX webpage, and in that webpage list
%also all the talks at conferences (Trento ECT*, SPIN2004, etc...)   
%This note summaries the outcome of the PAX work, anticipating
%the main lines of the emerging proposal, the physical motivations and the
%most important expected results. This might help your evaluation of the
%final complete project which will be presented in January.
%Still, achieving this ambitious goal requires crossing uncharted
%terrain and calls for the staged approach 
%We propose a staged approach in which all the
%major itemsf can be tested and optimized before going 
%to the most challenging part of the project: the polarized
%proton-antiproton collider. 
\begin{itemize}
\item Polarization buildup in the HESR ring, operated at the lowest
  possible energy, as discussed in PAX LoI, does not allow one to
  achieve the optimum degree of polarization in the antiproton beam.
  The goal of achieving the highest possible polarization of
  antiprotons and optimization of the figure of merit dictates that
  one polarizes antiprotons in a dedicated low--energy ring. The
  transfer of polarized low--energy antiprotons into the HESR ring
  requires pre--acceleration to about 1.5~GeV/c in a dedicated booster
  ring.  Simultaneously, the incorporation of this booster ring into
  the HESR complex opens up, quite naturally, the possibility of
  building an asymmetric antiproton--proton collider\footnote{It
    should be noted that within the PAX collaboration we realized the
    possibility of building an asymmetric collider only later, i.e.
    after the oral presentation at SPIN2004.}.
\end{itemize}
The PAX collaboration proposes an approach that is composed of two
phases. During these the major milestones of the project can be tested
and optimized before the final goal is approached: An asymmetric
proton--antiproton collider, in which polarized protons with momenta
of about 3.5 GeV/c collide with polarized antiprotons with momenta up
to 15 GeV/c. These circulate in the HESR, which has already been
approved and will serve the PANDA experiment. In the following, we
will briefly describe the overall machine setup of the HESR complex,
schematically depicted in Fig.~\ref{CSRring}.

Let us summarize the main features of the accelerator setup:
\begin{itemize}
\item[1.] An Antiproton Polarizer (AP) built inside the HESR area with
  the crucial goal of polarizing antiprotons at kinetic energies
  around $\approx 50$~MeV ($p\approx 300 $ MeV/c), to be accelerated
  and injected into the other rings.
  
\item[2.] A second Cooler Synchrotron Ring (CSR, COSY--like) in which
  protons or antiprotons can be stored with a momentum up to 3.5
  GeV/c.  This ring shall have a straight section, where a PAX
  detector could be installed, running parallel to the experimental
  straight section of HESR.
  
\item[3.] By deflection of the HESR beam into the straight section of
  the CSR, both the collider or the fixed--target mode become
  feasible.

\end{itemize}
It is worthwhile to stress that, through the employment of the CSR,
effectively a second interaction point is formed with minimum
interference with PANDA. The proposed solution opens the possibility
to run two different experiments at the same time.\\

In the following sections, we discuss the physics program, which
should be pursued in two different phases.

\subsection{Phase I}

A beam of unpolarized or polarized antiprotons with momentum up to 3.5
GeV/c in the CSR ring, colliding on a polarized hydrogen target in the
PAX detector. This phase is independent of the HESR performance.
  
This first phase, at moderately high energy, will allow for the first
time the measurement of the time--like proton form factors in single
and double polarized $\bar{p}p$ interactions in a wide kinematical
range, from close to threshold up to $Q^2=8.5$~GeV$^2$.  It would
enable to determine several double spin asymmetries in elastic
$\bar{p}^{\uparrow}p^{\uparrow}$ scattering.  By detecting back
scattered antiprotons one can also explore hard scattering regions of
large $t$: In proton--proton scattering the same region of $t$
requires twice the energy.  There are no competing facilities at which
these topical issues can be addressed.  For the theoretical
background, see the PAX LoI \cite{paxloi} and the recent review paper
\cite{brodsky}.
  
\subsection{Phase II} 

This phase will allow the first ever direct measurement of the quark
transversity distribution $h_1$, by measuring the double transverse
spin asymmetry $\Att$ in Drell--Yan processes $p^{\uparrow}
\bar{p}^{\uparrow} \rightarrow e^+ e^- X$ as a function of Bjorken $x$
and $Q^2$ (= $M^2$)
$$\Att \equiv \frac{d\sigup-d\sigdw}{d\sigup+d\sigdw}\,=\,
\hat{a}_{TT}\frac{\sum_q e_q^2
  h_1^q(x_1,M^2)h_1^{\overline{q}}(x_2,M^2)} {\sum_q e_q^2
  q(x_1,M^2)\overline{q}(x_2,M^2)}\,,$$
where
$q=u,\overline{u},d,\overline{d}\ldots$, $M$ is the invariant mass of
the lepton pair and $\hat{a}_{TT}$, of the order of one, is the
calculable double--spin asymmetry of the QED elementary process
$q\overline{q}\rightarrow e^+ e^-$.  Two possible scenarios might be
foreseen to perform the measurement, which are discussed below.

\subsubsection{Asymmetric collider} 

A beam of polarized antiprotons from 1.5 GeV/c up to 15 GeV/c
circulating in the HESR, colliding on a beam of polarized protons with
momenta up to 3.5 GeV/c circulating in the CSR.  This scenario however
requires to demonstrate that a suitable luminosity is reachable.
Deflection of the HESR beam to the PAX detector in the CSR is
necessary (see Fig.~\ref{CSRring}).
    
By proper variation of the energy of the two colliding beams, this
setup would allow a measurement of the transversity distribution $h_1$
in the valence region of $x>0.05$, with corresponding $Q^2=4 \ldots
100$ $\rm GeV^2$ (see Fig.~\ref{Figphysics}). $\Att$ is predicted to
be larger than 0.3 over the full kinematic range, up to the highest
reachable center--of--mass energy of $\sqrt{s}\sim\sqrt{200}$. The
cross section is large as well: With a luminosity of $5\cdot
10^{30}$~cm$^{-2}s^{-1}$ about $2000$ events per day can be
expected\footnote{A first estimate indicates that in the collider mode
  luminosities in excess of $10^{30}$~cm$^{-2}$s${^-1}$ could be
  reached. We are presently evaluating the influence of intra--beam
  scattering, which seems to be one of the limiting factors.}. For the
transversity distribution $h_1$, such an experiment can be considered
as the analogue of polarized DIS for the determination of the helicity
structure function $g_1$, i.e. of the helicity distribution $\Delta
q(x,Q^2)$; the kinematical coverage $(x,Q^2)$ will be similar to that
of the HERMES experiment.
    
\subsubsection{High luminosity fixed target experiment} 

If the required luminosity in the collider mode is not achievable, a
fixed target experiment can be conducted.  A beam of 22 GeV/c (15
GeV/c) polarized antiprotons circulating in the HESR is used to
collide with a polarized internal hydrogen target.  Also this scenario
requires the deflection of the HESR beam to the PAX detector in the
CSR (see Fig.~\ref{CSRring}).
  
\begin{figure}[hbt]
  \includegraphics[width=0.49\linewidth]{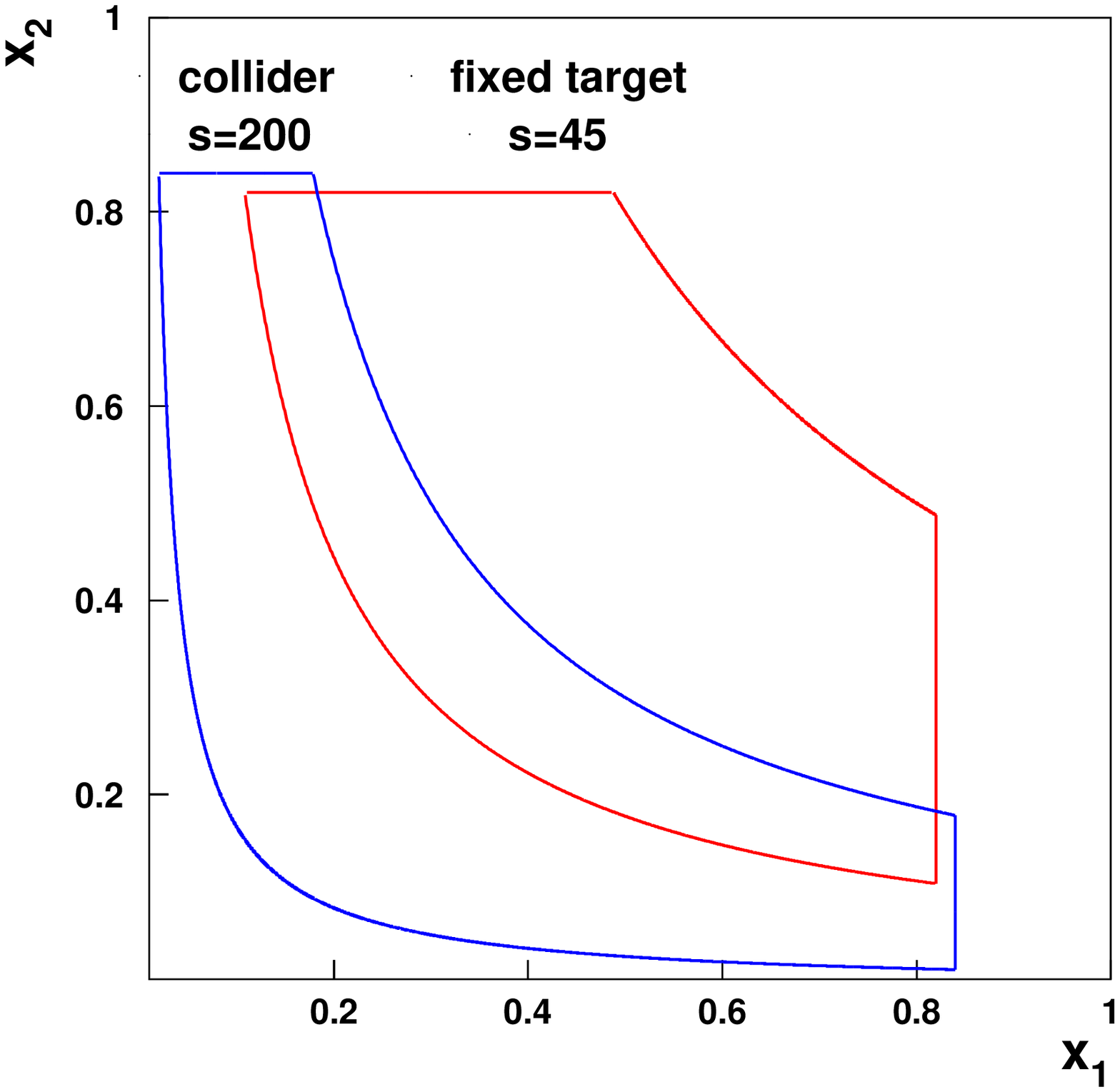}
  \includegraphics[width=0.49\linewidth]{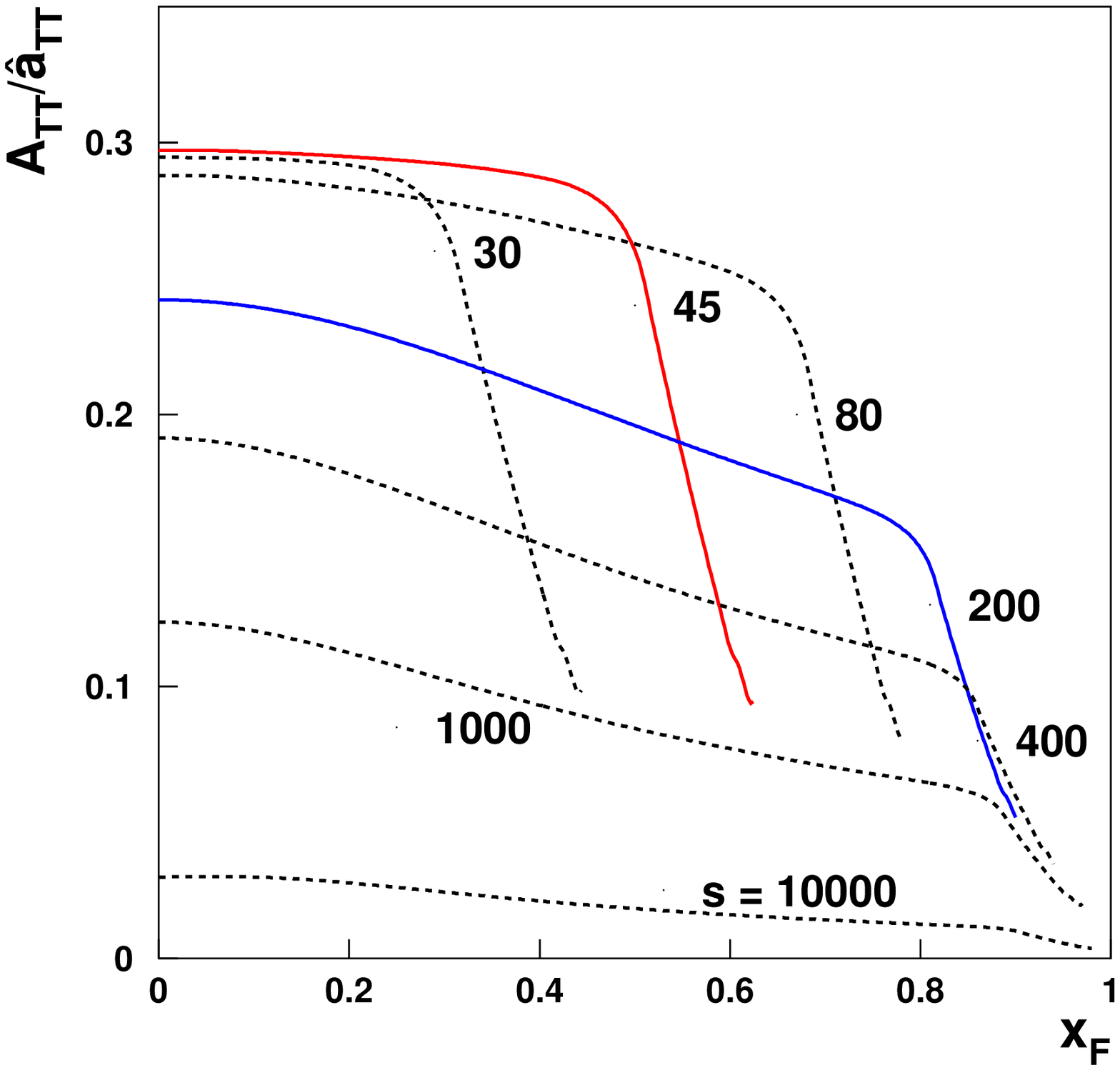}
\caption{\small Left: The kinematic region covered by the $h_1$ measurement 
  at PAX in phase II. In the asymmetric collider scenario antiprotons
  of 15~GeV/c impinge on protons of 3.5~GeV/c at c.m.  energies of
  $\sqrt{s}\sim \sqrt{200}$~GeV and $Q^2>4$ $\rm GeV^2$.  The fixed
  target case represents antiprotons of 22~GeV/c colliding with a
  fixed polarized target ($\sqrt{s}\sim\sqrt{45}$~GeV).  Right: The
  expected asymmetry as a function of Feynman $x_F$ for different
  values of $s$ and $Q^2=16$ $\rm GeV^2$.}
\label{Figphysics}
\end{figure}

A theoretical discussion of the significance of the measurement of
$\Att$ for a 22~GeV/c (15~GeV/c) beam impinging on a fixed target is
given in Refs. \cite{jpsi,efremov,zavada} and the recent review paper
\cite{brodsky}. The theoretical work on the $K$--factors for the
transversity determination is in progress \cite{Ratcliffe,barone}.
This measurement will explore the valence region of $x>0.2$, with
corresponding $Q^2=4 \ldots 16$ ${\rm GeV}^2$ (see
Fig.~\ref{Figphysics}).  In this region $\Att$ is predicted to be
large (of the order of 0.3, or more) and the expected number of
events can be of the order of 2000 per day.

We would like to mention, that we are also investigating whether the
PANDA detector, properly modified, is compatible with the transversity
measurements in the collider mode, where an efficient identification
of the Drell--Yan pairs is required. At the interaction point, the
spins of the colliding protons and antiprotons should be vertical,
with no significant component along the beam direction.

\section{Conclusion}
To summarize, we note that the storage of polarized antiprotons at
HESR will open unique possibilities to test QCD in hitherto unexplored
domains.  This will provide another cornerstone to the antiproton
program at FAIR.

\section*{Acknowledgments}
The author would like to especially thank Paolo Lenisa for his
contribution to the PAX project. In addition, the help of M.
Anselmino, D.  Chiladze, M.  Contalbrigo, P.F.  Dalpiaz, E. De
Sanctis, A.  Drago, A.  Kacharava, A.  Lehrach, B.  Lorentz, G.
Macharashvili, R.  Maier, S.  Martin, C.  Montag, N.N.  Nikolaev, E.
Steffens, D.  Prasuhn, H.  Str\"oher, and S.  Yaschenko is gratefully
acknowledged.

\end{document}